# The most common habitable planets - atmospheric characterization of the subgroup of fast rotators

## R. Pinotti [1,2]

1 – Observatório do Valongo, Universidade Federal do Rio de Janeiro - UFRJ, Ladeira Pedro Antônio 43, 20080-090, Rio de Janeiro, RJ, Brasil
2 - Petrobras, Av. República do Chile, 65, 21th floor, AB-RE/TR/PR, Rio de Janeiro RJ, 20031-912, Brasil

**ABSTRACT**

The current search for habitable planets has focused on Earth-like conditions of mass, volatile content and orbit. However, rocky planets following eccentric orbits, and drier than the Earth, may be a more common phenomenon in the Universe. For the subgroup of fast rotators, it is suggested that their atmospheric thermal capacitance, subject to the radiative forcing of their parent stars, may provide researchers in the near future with a simple method for the determination of a robust lower limit of atmospheric thickness. This technique, together with the  spectroscopic analysis of resolved planets from their stars, both allowed by planned space and ground-based observatories with thermal IR capabilities, would enable us with a better understanding of the habitability of this class of planets. The technique works better for smaller orbital periods, but since the tidal lock radius of M dwarfs encompasses their HZ, the optimum targets would be planets around K dwarf stars. The atmospheric thermal capacitance could also expand the range of Habitable Zones for shorter orbits, particularly for planets around M dwarf stars, since the  higher frequency of the periodic radiative forcing dampens the surface temperature variation considerably.

## 1  INTRODUCTION

The search for life elsewhere in the Cosmos has been one of the most fascinating pursuits of science, and so far a fruitless one, despite decades of research in many areas, from the study of radio signals that might reveal the existence of intelligent life (Tarter et al. 2010, Davies 2010), to the analysis of meteorites from Mars that might reveal fossilized microorganisms (McKay et al. 1996, Scott, Yamaguchi & Krot 1997).

As for the existence of habitable worlds around other stars, the discovery of hundreds of planets orbiting solar-like stars since 1995 has provided the field of astrobiology with a wealth of information, including the probability of a star harboring a planet, something that was a mere speculation when Drake created his famous formula (Glade, Ballet & Bastien 2012) for the number of advanced civilizations in the Galaxy, half a century ago. Most of the planets discovered so far, mainly by the Doppler-shift and the transit methods (Schneider 2012, Schneider et al. 2011, Wright et al. 2011), are giant planets, but an increasing number have masses of the same order of magnitude of the Earth's, and even smaller. Some of them orbit their stars within the so called Habitable Zone or HZ (Pepe et al. 2011, Wordsworth et al. 2011, Vogt et al. 2010), where water could be found in the liquid state. The possibility that these planets may harbor life could, in the near future, be strengthened or weakened, with the construction and deployment of space and ground-based observatories that will be able to analyze their atmospheric composition, looking for key molecules.

* Email: rpinotti@astro.ufrj.br

Exoplanets in general are expected to have, on average, more eccentric orbits than the planets around our solar system (Spiegel 2010, Spiegel 2010b, Kita, Rasio & Takeda 2010, Mandell, Raymond & Sigurdsson 2007). The variable stellar radiative flux will affect their atmosphere temperatures, and by observing the evolution of their brightness temperature it would be possible to learn about atmosphere thickness and, perhaps, about life.

Moreover, rocky planets around M dwarfs, which are by far the most numerous class of stars in our galaxy, are expected to be deficient in volatiles (Lissauer, 2007). As for rocky planets around main sequence stars in general, it is possible that low metallicity protoplanetary discs produce planets closer to their stars than do the ones with higher metallicity (Pinotti et al. 2005), and closer planets would have more difficulty in accreting volatiles. Therefore it is reasonable to suppose that rocky planets drier than the Earth, following eccentric orbits, are the majority concerning habitable worlds.

The dynamics of the brightness temperature of this class of rocky planets, considering the subgroup of fast rotators, will be explored in detail in the next sections, and results on atmospheric thickness and enhanced HZs around low mass stars will be discussed. The development of the model in the next section considers that the amount of liquid water present at the surface of these planets is not enough to affect the thermal inertia budget considerably. This constraint would not *a priori* rule out this class of planet from the classification of habitable.

In section 2 a model for the dynamics of the brightness temperature of a planet, subjected to the conditions discussed above, is developed. Section 3 shows computer simulation results, using the model, for three putative planets, each orbiting a main sequence star near the Sun, and highlights the prospects and constraints of the brightness temperature dynamics as a method for the estimation of atmospheric thickness. In section 4 the results are discussed, and a linearization of the main equation is developed in order to illustrate the importance of the orbital period on the brightness temperature oscillation.

## 2 MODELING THE DYNAMICS OF THE BRIGHTNESS TEMPERATURE FOR AN ECCENTRIC AND DRY ROCKY PLANET

The equilibrium temperature of a planet is a result of the radiative energy balance between the energy absorbed from the star and the energy emitted by the planet. This balance is mathematically translated, for a rapidly rotating planet (De Pater & Lissauer 2001), into Eq. (1)

$$F_{star}\pi R_p^2(1-A) = \varepsilon 4\pi\sigma R_p^2 T_p^4 \qquad (1)$$

where $F_{star}$ is the flux from the star integrated over the entire electromagnetic spectrum, A is the planet´s bond albedo, $R_p$ and $T_p$ are the planetary radius and equilibrium temperature respectively, and $\varepsilon$ is the emissivity, which is close to 0.9 at infrared. Here, with the objective of simplification, we consider that the planet is far enough from the star, and inside the HZ, so that no tidal lock phenomena has to be taken into account, and, finally, that the planet rotates fast enough and has an atmosphere thick enough so that the temperature on the night side does not drop significantly. More specifically, the radiation timescale, must be greater than the rotation period. As a consequence, the (thermal) radiation leaving the planet can be considered over the entire surface $4\pi R_p^2$.

Eq. (1) gives a straightforward and robust result for $T_p$, independently of planet radius, as a consequence of the match between the energy absorbed from sunlight and the energy radiated; exceptions in own solar system are Jupiter, Saturn and Neptune, due to the presence of an internal heat source, because their interiors are cooling or becoming more centrally condensed (De Pater & Lissauer 2001). As for terrestrial planets, especially Earth, Eq. (1) holds because the low orbital eccentricity guarantees a small variation of $F_{star}$ as the planets follow the path dictated by the keplerian laws, and because the atmosphere has a high thermal capacitance.

However, what about a hypothetical rocky planet with a more eccentric orbit? If $F_{star}$ oscillates substantially over the planetary orbit and the thermal capacitance of the atmosphere of the planet is not high enough, the radiative equilibrium will probably not hold, especially in the region near periastron, where the value of $dF_{star}/dt$ is highest.

A second look at Eq. (1), considering now that thermal equilibrium is not achieved, gives:

$$F_{star}\pi R_p^2(1-A) - \varepsilon 4\pi\sigma R_p^2 T_B^4 = C\frac{dT_B}{dt} \qquad (2)$$

Since Eq. (2) involves a time derivative of T, the term equilibrium is no longer quite accurate, therefore, in order to make it clear, I changed the subscript of T from P to B, meaning that from now on the temperature considered is the brightness temperature; $CdT_B/dt$ is the accumulation term and $C$ is the thermal capacitance of the atmosphere. Considering an oscillatory nature of $F_{star}$, now a periodic *forcing*, and a negligible value of $C$, that is, no significant atmospheric thermal capacitance, $T_B$ will oscillate *in phase* with $F_{star}$ according to Eq. (1). In other words, rocky planets with very thin or no atmosphere would achieve immediate thermal equilibrium, and $T_B$ would evolve in lockstep with the evolution of $F_{star}$.

On the other hand, Eq. (2) shows that for a high enough value of C, $dT_B/dt$ is negligible, that is, the *damping* caused by the large value of $C$ prevents the surface temperature from drifting noticeably away from its average value.

A planet under the influence of a significant variation of $F_{star}$ over its orbit, and with a not too high radiative atmospheric capacitance would exhibit both damping and phase lag of $T_B$ (Spiegel, Menou & Scharf 2008) relative to the situation with $C = 0 \ J \ K^{-1}$ (no atmosphere). If we could track the planet´s brightness temperature along its orbit, the resulting profile could be compared to the theoretical $C=0 \ J \ K^{-1}$ profile, thus providing us with valuable information on C and A.

The capacitance C can be modeled as

$$C = nc_p$$

where $n$ is the total mole number of the gas in the region of the atmosphere where heat transfer is mainly radiative, and $c_p$ is the mole heat capacity of the gas.

An estimate of $n$ can be made through Eq. (3):

$$n = 4\pi R_p^2 h\rho \qquad (3)$$

where $\rho$ is the mean mole density and h is the thickness of the atmosphere, both for the region where radiative heat transfer is dominant. With Eq. (3) and Eq. (2), the following equation is derived:

$$F_{star}\frac{(1-A)}{4} - \varepsilon\sigma T_B^4 = c_p h\rho \frac{dT_B}{dt}$$

Since the term $h\rho$ is a measure of the number of absorbers/emitters per area, or a column density, it is more appropriately defined as

$$h\rho = \chi$$

Then, we arrive at Eq. (4):

$$F_{star}\frac{(1-A)}{4} - \varepsilon\sigma T_B^4 = \chi c_p \frac{dT_B}{dt} \qquad (4)$$

It is important to notice that Eq. (4) is independent of $R$, so it could be used along the mass range from sub-Earth to super-Earth. One should keep in mind that, since the deposition depth varies from planet to planet, Eq. (4) would account for a lower limit of atmosphere thickness.

Eq. (4) depends on both $\chi$ and $A$. So, for a sufficient number of observations of the planet´s brightness temperature along its orbit, we can evaluate these two important parameters. The dynamical model depicted in Eq. (4) can also take into account the more realistic possibility that $A = A(T_B)$, as would be expected for planets with volatiles which produces cloud cover and/or planets with biological activity which depends strongly on temperature. However, in this work $A$ will be considered constant.

Here again we remember that, for Eq. (4) to be of use, the radiative forcing should be relevant, that is, the orbital eccentricity should be significant, so that variations in brightness temperature are not negligible.

It is important to stress that the atmosphere models used for simulations of thermal phase curves of tidally-locked rocky planets on circular orbits (Selsis, Wordsworth & Forget 2011, Maurin et al. 2012) are not, in principle, applicable to the scenario of long and eccentric orbits, since the planets would not be tidally locked (and the rotation would consequently dilute the night-day contrast in thermal emission), and the stellar radiative forcing would have a magnitude that renders any spatial inhomogeneities in the atmosphere negligible, when compared with the response of the whole planet´s brightness temperature (the planet considered as a point source).

As for the contribution of the solid surface of the planet to the thermal balance, the low thermal inertia of the various possible materials (Putzig & Mellon 2007) do not measure up with the thermal capacity of a substantial atmosphere. In the case of Mars, for example, there is around 174 kg of atmospheric $CO_2$ for each surface square meter, which means a potential column heat capacitance[1] of $1.27\times10^5$ J $K^{-1}$ $m^{-2}$. This guarantees that a near instantaneous thermal equilibrium of the soil is achieved with stellar radiation (Spencer 1990), leaving the thermal dynamics of the solid surface decoupled from the atmospheric one.

Since the thermal capacitance of the atmosphere becomes more evident when $dF_{star}/dt$ is greater, we should look for its effects near periastron. In the next section some results from computer simulations are shown in order to illustrate the potentials and limitations of the technique.

## 3 SIMULATING THE SCENARIO

In order to solve Eq. (4) for a number of cases, I developed a program for the calculation of the periodic radiative forcing, that is, a function that calculates $F_{star}$ as the planet follows its keplerian orbit. The program allows the choice of the star´s effective temperature, radius and mass, plus the semi-major axis and eccentricity of the planet´s orbit.

Then the program was coupled to the solver DASSLC (Differential-Algebraic System Solver in C) adapted for use in Matlab (Secchi 2010). The solver integrates Eq. (4) for given values of bond albedo, emissivity (considered constant and equal to 0.9) and column density ($\chi$).

In this study the heat capacity $c_p$ is considered to be constant ($c_p = 32$ J $mol^{-1}$ $K^{-1}$) and independent of the chemical composition of the atmosphere. This is a reasonable assumption, if one considers the range of temperatures in or near the HZ ($\sim 300$ K) and the susceptibility of heat capacity with temperature for gases relevant to rocky planet´s atmospheres (see table 1).

**Table 1.** Heat Capacities of Organic and Inorganic Compounds in the Ideal Gas State; temperature is expressed in Kelvin; $c_p$ is expressed in J $mol^{-1}$ $K^{-1}$

| Name | T | $c_p$ | T | $c_p$ | T | $c_p$ |
|---|---|---|---|---|---|---|
| $CH_4$ | 50 | 33.3 | 300 | 35.8 | 1500 | 88.9 |
| $O_2$ | 50 | 29.1 | 300 | 29.4 | 1500 | 36.5 |
| $N_2$ | 50 | 29.1 | 300 | 29.1 | 1500 | 34.8 |
| $NH_3$ | 100 | 33.4 | 300 | 35.6 | 1500 | 66.5 |
| $H_2O$ | 100 | 33.4 | 300 | 33.6 | 2273.15 | 52.8 |
| $CO_2$ | 50 | 29.4 | 300 | 37.3 | 5000 | 63.4 |
| $H_2S$ | 100 | 33.3 | 300 | 34.2 | 1500 | 51.4 |

Source: Perry (1997)

Since $F_{star}$ is particular to a given star and orbit, I selected three nearby stars, each a representative of G, K and M dwarf stars, for a set of simulations with habitable worlds. These putative planets (PP's) will have a range of values for eccentricity and $\chi$.

[1] *(174 kg $CO_2$ $m^{-2}$)*(32 J $mol^{-1}$ $K^{-1}$)*(22.73 mol $kg^{-1}$)*

### 3.1 Case Study: PP01 orbiting Tau Ceti

Table 2 lists the properties of the stars used in the case studies. Tau Ceti is close to the Sun, less massive but still massive enough so that the HZ will not be too close to the star, avoiding the possibility of tidal lock, according to Kasting, Whitmire & Reynolds (1993) who calculated the tidal lock radius for earth-sized planets in a 4.5 Gyr system with Q = 100 and initial rotation period of 13 h; the distance of the HZ also prevents a dangerous proximity to a potentially active stellar surface. However, as an old star, Tau Ceti offers little risk as chromospheric activity goes. Moreover, Tau Ceti seems to be rotating nearly face-on relative to the Solar System (Gray & Baliunas 1994), which would ease the determination and tracking of the orbit of our PP01. The very low metallicity of Tau Ceti of - 0.50 (Porto de Mello, del Peloso & Ghezzi 2006, Flynn & Morell 1997) indicates a lower possibility of it harboring large planets (Fisher & Valenti 2005, Gonzalez 1998). On the other hand, the extensive debris disk discovered (Greaves et al. 2004) indicates that it had the necessary material for rocky planet formation, and the lower metallicity may indicate that PP01 was formed closer to the star than was the case for our Solar System (Pinotti et al. 2005). Recent research on exoplanet candidates discovered by Kepler mission (Buchhave et al. 2012) also indicate that terrestrial planets do not require an enhanced metallicity environment in order to be formed.

The semi-major axis of the orbit was chosen to be 0.78 AU, which lies inside the Habitable Zone for this G8V star (Kasting, Whitmire & Reynolds 1993). This distance from the star translates to a maximum

**Table 2.** Properties of Tau Ceti, Gliese 687 and HD 219134

| Property | Tau Ceti | Gliese 687 | HD 219134 |
|---|---|---|---|
| Mass ($M_{sun}$) | 0.783 | 0.401 | 0.850 |
| Radius ($R_{sun}$) | 0.793 | 0.492 | 0.684 |
| Temperature (K) | 5,344 | 3,095 | 5,100 |
| Distance (ly) | 11.91 | 14.77 | 21.19 |

References: Teixeira et al. 2009, Santos et al. 2004, Berger et al. 2006, Porto de Mello, del Peloso & Ghezzi 2006

0.21 arcsec separation from Tau Ceti, as observed from Earth. The planet's bond albedo was chosen to be 0.25, an intermediate value between the Earth's and Mars'.

Table 3 shows the simulation results for five different values of eccentricity and three values of $\chi$. The unit of $\chi$ is mol m$^{-2}$. In order to transform it to number column density (cm$^{-2}$) multiply by $6.02 \times 10^{19}$. The attenuation values are the differences of $T_B$, between a planet with $\chi$ =0 at periastron and the maximum value of $T_B$ for a planet with the corresponding values of $\chi$. The phase lag is the difference in time of the corresponding values of $T_B$. The second column at Table 3 gives also the value of $T_B$ at periastron with $\chi$ =0. Figure 1 illustrates the effect for $\chi = 10^5$ mol m$^{-2}$ and eccentricity = 0.100. Here it is noteworthy to stress that, although the values of $T_B$ in the simulation are always below the freezing point of water, they correspond to the radiation emission of the atmosphere to outer space; the temperature at the surface of the planet is expected to be higher due to the greenhouse effect, and its value depends on the atmospheric composition.

The data displayed on Table 3 reveals the potential of the method for $\chi$ on the order of $10^5$ mol/m$^2$, which is where the Earth is located, with $\chi = 3.4 \times 10^5$ mol m$^{-2}$, considering the total atmosphere, and roughly half that value for the fraction where heat transfer is predominantly radiative (De Pater & Lissauer 2001). The attenuation of $T_B$ at periastron ($\Delta T_B > 1\% \ T_B$) and the phase lag would be easily noticed by future observatories with thermal IR capabilities that could resolve the planet from the star.

For values of the order of $10^4$ mol m$^{-2}$. the attenuation becomes perhaps too small for detection and tracking, although for high values of orbital eccentricity (> 0.200) the values improve. As an example of the limitation for thin atmospheres, a simulation for the case of Mars, which has a high orbital eccentricity of nearly 0.1, gives a value of attenuation below 0.5 K. Another factor that contributes for this low value is the high value of the orbital period for Mars.

On the other extreme, values of $\chi$ on the order of $10^6$ mol m$^{-2}$ are almost certainly an unrealistic physical situation, since the phase lag becomes too great (~ $10^6$ s), one order of magnitude higher than the assumed radiative timescale for the radiative region of an extrasolar planet's atmosphere, as will be shown in the Discussion and Conclusions section. Still, there is a wide range of $\chi$ which would be useful for future researchers of rocky planets, habitable or not.

As for the effect of the selected range of eccentricities on habitability, it is possible that values of 0.3 or even higher do not rule out the possibility of life (Williams & Pollard 2002).

### 3.2 Case Study: PP02 orbiting Gliese 687

Rocky planets with sub-Earth mass are expected to be present in extrasolar planetary systems, especially around the more numerous M Dwarfs (Lépine & Gaidos 2011), in greater quantity than earth-sized planets (Montgomery & Laughlin 2009, Cassan et al. 2012). The fact that the M dwarf star KOI-961 has recently been found to harbor three planets smaller than the Earth (Muirhead et al. 2012) lends credence to this scenario. On the other hand, results from observational research (Delfosse et al. 2012, Bonfils et al. 2011) indicate that M dwarfs are also abundant with super-Earth planets. So, a case study with an M dwarf is likely to represent many, if not most, of the habitable planets.

In order to study the effectiveness of the method for the most common main sequence class of stars, I have chosen the nearby Gliese 687 as a representative of the M dwarf population of the solar neighborhood (see table 2). At a distance of 14.77 light-years, it is close enough for searches using future observatories with direct imaging capabilities. Its high metallicity (Berger et al. 2006) of + 0.11 indicates a good chance that rocky planets have formed around the star during the protoplanetary disc lifetime. Our putative planet orbiting Gliese 687 has the same bond albedo of the former simulation (0.25) and its semi-major axis is 0.17 AU, which puts it inside the HZ (Kasting, Whitmire & Reynolds 1993). Although any old planet (older than ~ 1 Gyr) in the HZ of an M dwarf is potentially subject to tidal lock, Gliese 687 is massive enough so that the tidal lock radius is very close to 0.17 AU, and I assume that our PP02 is not tidal locked, and that the equations developed in this paper are still valid.

Table 3 – Results of simulations for PP01 around Tau Ceti

| Eccentricity | $T_B$ at periastron with $\chi = 0$ | $\chi = 10^4$ mol m$^{-2}$ | | $\chi = 10^5$ mol m$^{-2}$ | | $\chi = 10^6$ mol m$^{-2}$ | |
|---|---|---|---|---|---|---|---|
| | | Phase lag (h) | Att. (K) | Phase lag (h) | Att. (K) | Phase lag (h) | Att. (K) |
| 0.050 | 254.66 | 15.48 | 0.64 | 152.25 | 2.80 | 892.95 | 7.82 |
| 0.075 | 258.08 | 14.93 | 0.97 | 148.60 | 4.27 | 888.90 | 11.89 |
| 0.100 | 261.64 | 14.38 | 1.31 | 144.90 | 5.79 | 873.35 | 15.98 |
| 0.200 | 277.50 | 12.23 | 2.80 | 129.67 | 12.42 | 811.58 | 33.21 |
| 0.300 | 296.67 | 10.18 | 4.52 | 113.73 | 20.18 | 774.70 | 53.27 |

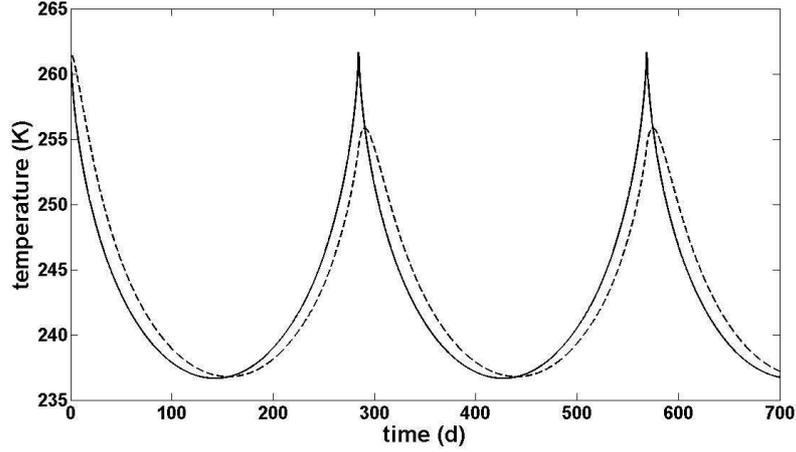

**Figure 1**. Results from the simulation of PP01 around Tau Ceti with $e$ = 0.100. Continuous line refers to $\chi$ = 0; dashed line refers to $\chi$ = 10$^5$ mol m$^{-2}$. Note that the higher value for the dashed line at time = 0 is the initial guess required for the integrator DASSLC

**Table 4.** Results of simulations for PP02 around Gliese 687

| Eccentricity | $T_B$ at periastron with $\chi = 0$ | $\chi = 10^4$ mol m$^{-2}$ | | $\chi = 10^5$ mol m$^{-2}$ | |
|---|---|---|---|---|---|
| | | Phase lag (h) | Att. (K) | Phase lag (h) | Att. (K) |
| 0.050 | 248.84 | 16.42 | 2.30 | 112.85 | 7.19 |
| 0.075 | 252.18 | 15.98 | 3.50 | 111.48 | 10.91 |
| 0.100 | 255.66 | 15.55 | 4.75 | 110.07 | 14.71 |
| 0.200 | 271.16 | 13.77 | 10.17 | 103.97 | 30.74 |
| 0.300 | 289.86 | 11.95 | 16.49 | 97.15 | 49.36 |

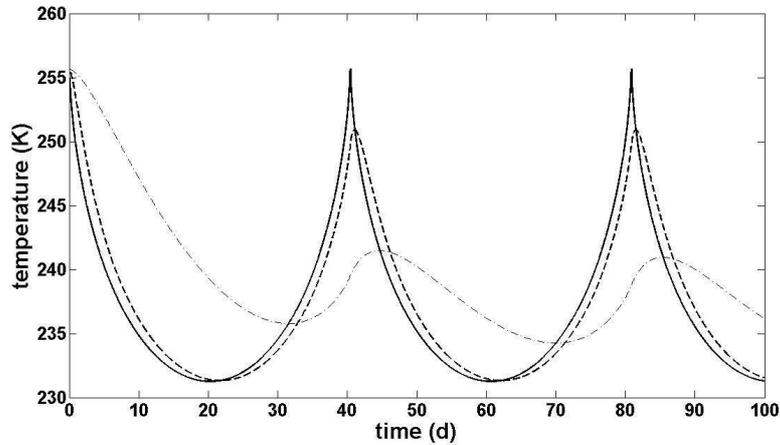

**Figure 2**. Results from the simulation of PP02 around Gliese 687 with $e$ = 0.100. Continuous line refers to $\chi$ = 0 mol/m$^2$; dashed line refers to $\chi$ = 10$^4$ mol m$^{-2}$ ; dashdot line refers to $\chi$ = 10$^5$ mol m$^{-2}$ Note that the higher values for the dashed and dashdot lines at time = 0 are the initial guess required for the integrator DASSLC

**Table 5.** Results of simulations for PP03 around HD 219134

| Eccentricity | $T_B$ at periastron with $\chi = 0$ | $\chi = 10^4$ mol m$^{-2}$ | | $\chi = 10^5$ mol m$^{-2}$ | |
|---|---|---|---|---|---|
| | | Phase lag (h) | Att. (K) | Phase lag (h) | Att. (K) |
| 0.050 | 253.17 | 15.77 | 0.83 | 152.88 | 3.54 |
| 0.075 | 256.56 | 15.22 | 1.26 | 149.78 | 5.40 |
| 0.100 | 260.11 | 14.68 | 1.70 | 146.63 | 7.32 |
| 0.200 | 275.86 | 12.55 | 3.62 | 133.38 | 15.72 |
| 0.300 | 294.88 | 10.48 | 5.82 | 119.10 | 25.55 |

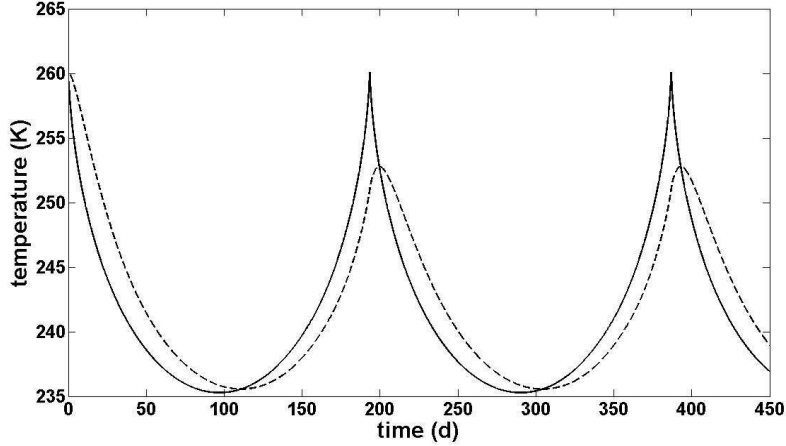

**Figure 3.** Results from the simulation of PP03 around HD 219134 with $e = 0.100$. Continuous line refers to $\chi = 0$; dashed line refers to $\chi = 10^5$ mol m$^{-2}$. Note that the higher value for the dashed line at time = 0 is the initial guess required for the integrator DASSLC

Table 4 shows the results for five values of eccentricity, and two values of $\chi$. The values of eccentricity are consistent with results of simulations of terrestrial planet formation around M dwarfs (Ogihara & Ida 2009). The data shows clearly that the attenuation values for $\chi = 10^4$ mol m$^{-2}$ and $\chi = 10^5$ mol m$^{-2}$ are considerably higher than the corresponding values for PP01 around Tau Ceti. The reason for this result is that the orbital period for PP02 is much shorter than that for PP01, that is, the frequency of the periodic radiative forcing has increased, increasing therefore the damping effect of the atmospheric heat capacitance. Figure 2 illustrates the increased damping effect for PP02, for eccentricity = 0.100.

The shorter orbital period for planets in the HZ of stars less luminous than Tau Ceti (K dwarfs and the most massive M dwarfs) means that the method could also be used to study planets with thin atmospheres. For less massive M dwarfs a correction factor should be used, in order to compensate for the tidal lock effect and the corresponding thermal phase variation that would become evident. It is also worth noting that for $\chi = 10^5$ mol m$^{-2}$ and eccentricity higher than 0.2, the values of attenuation of $T_B$ at periastron become substantial. ($\Delta T_B > 10\% \, T_B$).

### 3.3 Case Study: PP03 orbiting HD 219134

HD 219134 is a K3V dwarf both metal rich and near the Sun (Heiter and Luck 2003). Its high metallicity of +0.10 enhances the possibility of the presence of one or more planets, and the proximity of the star (6.5 pc) makes their detection easier. This star is also chromospherically inactive, which is another factor in favor of the habitability of its putative planets. As a consequence of these characteristics, HD 219134 has been regarded as a priority target for future space-based searches for habitable planets (Porto de Mello, del Peloso & Ghezzi 2006), along with Tau Ceti, among the stellar population within 10 pc of the Sun.

The mass, surface temperature and radius of HD 219134 (see Table 2) indicates that the selected location of PP03, at a semi-major axis of 0.62 AU, places it inside the HZ, and comfortably away from the distance where the tidal lock effect becomes significant (Kasting, Whitmire & Reynolds 1993). The fact that the mass of this K dwarf is slightly higher than that of the G dwarf Tau Ceti should not come as a surprise, since the metallicity difference between these stars is considerable

Table 5 and Figure 3 show the simulation results for the pair HD 219134/PP03. As expected, they are intermediate between the pair Tau Ceti/PP01 and the pair Gliese 687/PP02, although much closer to the former one. In the next section it will be shown quantitatively how the orbital period, much shorter in the case of Gliese 687, affects the damping effect.

## 4 DISCUSSION AND CONCLUSIONS

So far, all the planets detected by direct imaging belong to the class of gas giants (Neuhäuser & Schmidt 2012, Schneider 2012). Most of them orbit their host stars at a considerable distance and/or orbit young stars, so that the planets' luminosities are higher due to the energy release from gravitational potential of its material, still contracting. However, gas giants are not the focus of the technique presented in this paper. Rocky planets like the Earth are $10^9$ to $10^{10}$ dimmer, in the visible, than their host stars; even in the thermal IR, where the contrast is lower, the luminosity ratio is still on the order of $10^7$. This barrier is expected to be broken with the advent of the space telescope JWST (Charbonneau & Deming, 2007, Catanzarite & Shao 2011), with thermal IR capability, due for launch in 2018, and the ground-based E-ELT (Pantin, Salmon & Charnoz 2010, ESO 2011), capable of achieving a contras of 1:$10^9$ at 0.1 arcseconds, could observe super-Earth planets around the closest stars at thermal IR. Since the method described in this work does not put constraints on the planet´s radius, it would apply to hypothetical super-Earths in eccentric orbits that these two observatories might discover, provided that such planets have thin atmospheres ($10^4 < \chi < 10^5$ mol m$^{-2}$). Although super-earths are expected to have thick atmospheres, scaling with planetary radius (Porto de Mello, del Peloso and Ghezzi 2006), there may be exceptions for the ones in the HZ of M dwarfs, as it will be pointed out later in this section.

However, the technique described in this work would be of fullest use in planned space observatories such as TPF-I (Seager, Ford & Turner, 2002) and Darwin (Kaltenegger & Fridlund 2005). The SIM Planet Quest mission (Unwin et al. 2007, Goullioud et al. 2008, Catanzarite et al. 2006) would be an important player also, identifying terrestrial planets in the visible band and determining planet mass, albedo, radii and eccentricity. Although the projects for these observatories are currently cancelled, it is reasonable to expect that similar or identical projects will be completed in the near future (with the exception of SIM Planet Quest, as a result of recommendations from the Decadal Review), since they arguably represent the best concepts available with our current technology.

By using the equation of energy balance (2), researchers of habitable planets usually do not take into account the atmospheric heating contribution (Kane & Gelino 2011, 2012), that is, the time derivative of surface temperature is considered zero. When they do take into account the thermal capacity, coupled to stellar flux forcing, the focus is the investigation of habitability conditions (Spiegel, Menou & Scharf 2008, Dressing et al. 2010), and not the study of the column density of the atmosphere as a whole. Atmospheric thermal capacity is also taken into account in models that predict thermal phase variations for close-in gas giant planets (Cowan & Agol 2011), but in this case no useful information on the column density of their atmospheres is possible, since the atmosphere layer responsible for radiation emission is negligible next to the bulk of the atmosphere of the gas giants.

One possible drawback of this technique is related to the (desired) possibility that the surface of the planet contains an appreciable quantity of volatiles such as water. Its presence would alter the planetary albedo periodically through the intensification of cloud cover and/or ice formation; and, for a really large surface fraction covered by water, the energy balance equation (4) should include values for sensible and latent heat. However, these factors can be included in the model also, and reflection spectra data (Kitzmann et al. 2011) plus atmospheric composition obtained by spectroscopy should give enough information for the necessary adjustments. For Earth-like planets, which in the future may well prove to be a subset of the class of habitable planets, more sophisticated modeling would be required (Cowan, Voigt & Abbot 2012).

The technique would be more effective for higher values of eccentricity and, mainly, for lower values of semi-major axis. In order to appreciate mathematically the influence of these two variables, let´s take Eq. (4), which is highly non-linear due to the terms $F_{star}$ and $T_B^4$, and calculate a linear approximation for the case of a near circular orbit.

Using Taylor series on the term $\sigma \varepsilon T_B^4$ :

$$\sigma \varepsilon T_P^4 \cong \sigma \varepsilon [T_B(a)]^4 + 4\sigma \varepsilon [T_B(a)]^3 [T_B - T_B(a)] \quad (5)$$

Where $T_B(a)$ is the steady-state value of the temperature at a circular orbit with radii $a$.

The term $F_{star}$ could be represented by a sinusoidal forcing :

$$F_{star} \cong F(a) + F(a)\left[\frac{1}{(1-e)^2} - 1\right]\sin(\omega t) \quad (6)$$

Where $F(a)$ is the incident flux at distance $a$, $e$ is the eccentricity and $\omega$ is the orbital frequency, which, according to Kepler´s Third Law can be written as

$$\omega = \frac{2\pi}{T} = \frac{(m_* G)^{0.5}}{a^{1.5}}$$

where $T$ is the orbital period, and $m_*$ is the stellar mass.

Using Eqs (5) and (6) in (4), and noting that for a steady-state circular orbit $\sigma \varepsilon [T_B(a)]^4 = F(a)(1-A)/4$, the final result can be written as

$$\overline{F} - 4\sigma\varepsilon[T_B(a)]^3 \overline{T_B} = \chi c_p \frac{d\overline{T_B}}{dt} \qquad (7)$$

Where $\overline{F}$ and $\overline{T_B}$ are deviation variables calculated around the steady-state circular orbit, and given by

$$\overline{T_B} = T_B - T_B(a)$$

$$\overline{F} \cong \frac{(1-A)}{4} F(a) \left[ \frac{1}{(1-e)^2} - 1 \right] \sin(\omega t)$$

The linear and first order system represented by Eq. (7) has a time constant given by

$$\tau = \chi c_p \left[ 4\sigma\varepsilon(T_B(a))^3 \right]^{-1}$$

For $T_B(a) \cong 260$ K and $\chi = 10^5$ mol m$^{-2}$ the result is $\tau \cong 8.9$ x $10^5$ s. The system can be solved analytically (Skogestad and Postlethwaite 2005, Seborg, Edgard & Mellichamp 1989); for a sinusoidal forcing $\overline{F} = \Psi\sin(\omega t)$ the response is also sinusoidal:

$$\overline{T_p} = K\Psi(\omega^2\tau^2 + 1)^{-0.5} \sin(\omega t + \phi) \qquad (8)$$

Where K is the system gain, in this case $K = \left[ 4\sigma\varepsilon(T_B(a))^3 \right]^{-1}$, and $\phi$ is the phase lag $\phi = -\tan^{-1}(\omega\tau)$. Eq. (8) shows that the amplitude of $\overline{T_B}$ is a function of $\omega$. More importantly, the amplitude ratio between $\overline{T_B}$ and $\overline{F}$ is

$$AR = \frac{K}{\sqrt{\omega^2\tau^2 + 1}}$$

The damping ratio, that is, the ratio between the AR and AR for the case of no atmosphere $(\tau = 0)$ is then

$$DR = \frac{1}{\sqrt{\omega^2\tau^2 + 1}} \qquad (9)$$

Eq (9) shows us not only that for increasing values of $n$ the damping ratio gets lower, but also that, for $\omega \gg 1/\tau$

$$DR \cong \frac{1}{\tau}\omega^{-1} \cong \frac{1}{\tau(Gm*)^{0.5}} a^{1.5} \qquad (10)$$

Eqs (9) and (10) show why the semi-major axis of the orbit is so influential on the damping ratio. Figure 4 illustrates the impact of $\omega$ on the value of DR for the case of $\tau = 8.9$ x $10^5$ s. For the case of the pair Gliese 687/PP02, $\omega = 1.801$x$10^{-6}$ rad s$^{-1}$, which, compared to the value of $1/\tau = 1.124$x$10^{-6}$, explains the large damping ratio found in the simulation.

The linear approximation informs us that the eccentricity is important for the amplitude of the forcing ($\Psi$), so that, for a given DR, higher eccentricities correspond to higher values of $\Psi$, which would ease the detection of the damping effect. High values of eccentricities are in principle a problem for habitability due to the intense surface temperature variation, but climate models (Williams & Pollard 2002) suggest that such planets could still be habitable, for the main factor for long-term climate stability would be the average stellar flux received over an entire orbit. The effect of a high obliquity could compensate for high eccentricity (Dressing et al. 2010), keeping the planet habitable out to greater distances. Also, in regions of the planet where there are large bodies of water, the extra thermal stability would enhance the prospects for habitability.

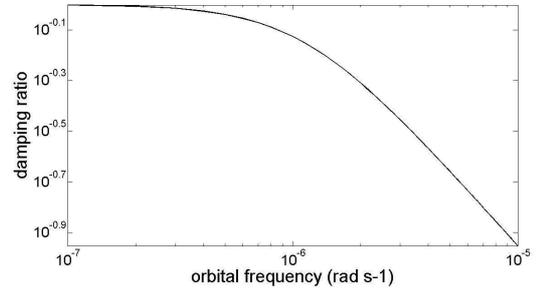

**Figure 4.** log ( $\omega$ ) x log (DR) for $\tau = 8.9$ x $10^5$ s

Very low values of semi-major axis give rise to potential tidal lock effects and vulnerability to stellar flares, more common on M dwarfs. So, it is unavoidable to foresee that some of the best data from the technique would come from planets where life is absent, or at least under a considerable pressure from its environment.

In the case of M dwarf stars it is possible that the quantity of volatiles available at the surface of planets inside or near the HZ is not enough to alter the bond albedo (Lissauer 2007) along their eccentric orbits, and the method depicted in this work could be used to assess this possibility also. Moreover, atmospheric erosion by strong winds from M dwarfs may impose an upper constraint on planetary atmospheric thickness and affect their habitability (Lammer et al. 2007). On the other hand, these factors raise the possibility that super-earths around M dwarfs could have atmospheres equivalent or even thinner than the Earth's ($\chi \leq 10^5$ mol m$^{-2}$), extending the mass range of the method. Moreover, for more massive M dwarf stars younger than ~ 1 Gyr, the tidal lock effect may be sufficiently small so that the constraint of fast rotation is no longer valid. Still, M dwarfs are likely to continue as important target stars for research on habitable planets (Tarter 2007).

For G dwarf stars the distance of the HZ region is considerably higher than that of M dwarf stars, eliminating the problems of tidal lock and vulnerability to flares. However, the high values of semi-major axis could impose a detection constraint on the method (see Table 6), with damping ratios higher than 0.95, although for high eccentricities, and consequently for the non-linear case of Eq. (4), the damping ratios increase substantially. On the other hand, M dwarfs present very low periods for planets in the HZ, and consequently the damping ratio becomes very attractive for the method, reaching values of around 0.5.

All these considerations lead us to conclude that, in terms of habitable worlds, the optimum target for the method would be K dwarf stars, which are very common and much longer lived than G dwarfs. For example, the value of DR for a planet with $\tau = 8.9 \times 10^5$ s , orbiting a 0.6 $M_{sun}$ K dwarf at 0.5 AU at low eccentricity would be 0.932, and even lower for more eccentric orbits, where linear approximation is no longer valid.

**Table 6.** Damping ratio for selected Planet/Star pairs, ranging from spectral type G to M, for the linear approximation of Eq. (7). The time constant $\tau \cong 8.9 \times 10^5$ s was used in the calculations

| Star/Planet | Semi-major axis (AU) | $\omega$ ($10^{-6}$ rad s$^{-1}$) | Period (d) | DR |
|---|---|---|---|---|
| Sun/Earth | 1.00 | 0.199 | 365 | 0.985 |
| Tau Ceti/ PP01 | 0.78 | 0.256 | 284 | 0.975 |
| HD219134/PP03 | 0.62 | 0.377 | 193 | 0.948 |
| Gliese 687/PP02 | 0.17 | 1.801 | 40 | 0.529 |

Another potential increase in the complexity of the brightness temperature is the possibility that life is present in a planet with eccentric orbit and that its surface would exhibit albedo variation due to the evolution of biological activity coupled with the stellar radiative flux oscillation; the dynamics of the total IR flux from the planet would also exhibit a different behavior compared with that of a merely habitable planet; we can imagine that vegetation, for example, would try to attenuate the temperature variation induced from to the large amplitude of stellar radiative flux, by albedo compensation (Watson & Lovelock 1983). In this case, spectroscopy would also be of great help in order to understand and model the phenomenon.

The fact that the attenuation of $T_B$ is more pronounced for shorter values of orbital period could in fact expand the range of the HZ. The simulation for Gliese 687 shows that, for example, the amplitude of $T_B$ for a planet at 0.17 AU and eccentricity = 0.100 drops from 24.4 K to 6.7 K for $\chi = 0$ and $\chi = 10^5$ mol m$^{-2}$ respectively. The prospects for habitability of a planet are greatly enhanced for steadier values of $T_B$, for many reasons. One is the potential for a runaway greenhouse effect caused by volatiles like water, which could render the surface of the planet uninhabitable when $T_B$ is too high. On the other hand, when $T_B$ becomes too low the ice-albedo positive feedback could render the surface gelid.

Although the effect of $T_B$ amplitude damping due to increasing values of $\omega$ has been mentioned in previous studies (Spiegel et al. 2010), this work shows how it can be used as a systematic method for atmosphere research. The damping can expand the HZ in two ways. Firstly, the thermal capacitance could make habitable a previously considered uninhabitable planet that departs from the standard HZ for some time due to its eccentric orbit; and secondly, it allows habitability for a planet with eccentric orbit which stays within the HZ, but has previously been considered uninhabitable because of the presumed high variation of $T_B$.

The fraction of the atmosphere that accounts for $\chi$ was not subject to analysis in this work, and theorists of exoplanet atmospheres would be able to set constraints with the help of models. For high values of eccentricity, for example, it is reasonable to suppose that the convective and radiative layers of the atmosphere could be periodically altered, even disrupted, perhaps increasing the radiative layers and therefore expanding the lower limit of atmospheric thickness provided by the technique.

The derivation of atmospheric column densities, from future observations of the dynamics of brightness temperatures of a variety of rocky planets with sub to super-Earth masses in eccentric orbits, is likely to become an important tool for the estimation of atmosphere extent, which, coupled with spectroscopic analysis, could bring us a clearer picture of atmospheric characteristics. And the effect of the thermal capacitance of their atmospheres on the stabilization of $T_B$ should be taken into account in the study of Habitable Zones, particularly for M dwarf stars.

# 5 ACKNOWLEDGMENTS

I would like to thank G. F. Porto de Mello for important suggestions on astrobiology and stellar characteristics, H. M. Boechat-Roberty for discussions and encouragement, D. S. Spiegel for an important review of the paper, A. R. Secchi, E. Almeida Neto and A. D. Quelhas for assintance on DASSLC and Matlab, and Solange Lins Klein.